\documentclass[11pt]{article}
\usepackage{moriond,epsfig}
\usepackage[latin1]{inputenc}
\usepackage[OT1]{fontenc}

\def\be{\begin{equation}}
\def\ee{\end{equation}}
\def\bea{\begin{eqnarray}}
\def\eea{\end{eqnarray}}

\newcommand{\rhalf}{$r_{1/2}$\,}
\newcommand{\rhalfRc}{($r_{1/2}, {\rm{R}}$)\,}
\newcommand{\Rc}{R\,}
\begin{document}
\vspace*{4cm}
\title{Mapping Dark Matter in Galaxy Clusters:\\
Gravitational Lensing \& Numerical Simulations}

\author{Marceau Limousin}

\address{
Laboratoire d'Astrophysique de Toulouse-Tarbes, Universit\'e de Toulouse, CNRS\\
57 avenue d'Azereix, 65\,000 Tarbes, France \\
Dark Cosmology Centre, Niels Bohr Institute, University of Copenhagen\\
Juliane Marie Vej 30, 2100 Copenhagen, Denmark
}

\maketitle\abstracts{
The different regimes of gravitational lensing constitutes an interesting tool in order
to map the mass distribution in galaxy clusters on different scales.
%The observational results can be compared to the predictions of numerical N-body
%simulations.
In this proceedings article, I review some work I have performed on this topic.
More precisely, I will focus on : (i) galaxy scale substructures, using weak 
galaxy-galaxy lensing in order to study
how does the environment shape their properties; (ii) the mass profile of Abell~1689
as probed combining strong and weak lensing; (iii) the slope of the inner dark matter 
distribution in Abell~1703 as measured by strong lensing.
The lensing results will be compared to the expectations from numerical simulations, 
when available.
}

\section{Gravitational Lensing by a Galaxy Cluster}
In this article, I will assume that the Newtonian dynamics still holds on large scales and will
interpret the results within the framework of dark matter, but note that viable alternative theories 
of gravity do exist.

The emerging picture coming from observational probes and numerical simulations is the following:
clusters of galaxies consist of 1-2\% stars; 10-20\% intracluster gas with a temperature of
1-15\,keV emitting in \textsc{x}-ray; 80-90\% dark matter.
The lensing effect, i.e. the bending of light by matter along the line of sight from the source
to the observer, depends only on the mass distribution of the intervening structures, making
gravitational lensing an interesting tool for measuring the mass profiles of lensing
structures. In particular, no additional assumptions need to be made
with regard to the dynamical state (relaxed or not, in hydrostatic
equilibrium or not) or the nature (baryonic or not, luminous or dark)
of the intervening matter. However, \emph{all} mass distributions
along the line of sight contribute to the lensing signal, introducing
contamination by foreground or background objects. In the core of
massive clusters, the surface mass density is well above the critical
value enabling the use of detected strong lensing features to
constrain the inner part of the cluster potential. At larger
cluster-centric radii, the ellipticities of weakly sheared background
galaxies is used to estimate the weak lensing effects induced by the
cluster potential.
Combining strong and weak lensing techniques allows one to map the mass distribution
of a galaxy cluster from the centre to the outskirts.
Moreover, galaxy scale substructures induce local modulation of the shear field in their neighbourhood.
This weak signal can be statistically detected in order to probe
the mass distribution of galaxy scale substructures.

\section{Truncation of Galaxy Dark Matter haloes in High Density Environment}

Galaxy scale dark matter haloes can be probed via a technique called 
weak galaxy-galaxy lensing.
The idea is the following : observing the shapes of distant background galaxies which have been 
lensed by foreground galaxies allows us to map the mass distribution of the foreground galaxies.
Of course, the lensing effect is small compared to the intrinsic ellipticity distribution of galaxies, 
thus a statistical approach is needed. Consequently, galaxy-galaxy lensing studies allow to constrain 
the mean properties of the halo population as a whole, and the reliability of the signal will depend on 
the number of foreground-background pairs. One advantage of this statistical approach is that it provides 
a probe of the gravitational potential of the halos of galaxies out to very large radii, where few 
classical methods are
viable, since dynamical and hydrodynamical tracers of the potential cannot be found at this radii.

\subsection{Theoretical Approach}
My first approach of galaxy-galaxy lensing has been theoretical (Limousin et~al., 2005 \cite{mypaperI}): 
I have been simulating data sets,
for different observational configurations and showed that a maximum likelihood method 
(Schneider \& Rix, 1997 \cite{rix}) is well adapted
to the problem, in the sense that it allows to retrieve the structural parameters that characterize 
galaxies' dark matter haloes: the central velocity dispersion, $\sigma_0$, which is related to the
depth of the potential well, and $r_{1/2}$, the half mass radius, which is related to the spatial
extent of the halo. Given this parametrization, the mass of the halo scales as M$\sim\,\sigma_0\,r_{1/2}$.
After extensive simulation work, the technique developed theoretically was applied
on a sample of five massive galaxy clusters at $z\sim0.2$.

\subsection{Application to a Sample of Cluster Lenses at $z\sim0.2$}
Data were taken at the \textsc{cfht} with the \textsc{cfh12k} camera through the B, R and I filters
(Czoske et~al.,  2002 \cite{olliphd}).
The object detection is described in Bardeau et~al. (2005) \cite{bardeau05}.
We selected the elliptical cluster galaxies as lenses.
We used only objects detected in all three bands and with reliable shape information, and
we undertook a Bayesian photometric study to derive a redshift estimation for each background galaxy.
Comparison to the colour-cut adopted in the \textsc{deep2} redshift survey (Coil et~al., 2004 \cite{deep2})
to select galaxies at $z>0.7$ showed that our procedure allowed to reliably discriminate
background galaxies.

The smooth component corresponding to the cluster itself was also considered.
The maximum likelihood method was applied on the galaxy catalogs with the goal to derive
some constraints averaged on the cluster galaxy population.
The main result of this analysis (Limousin et~al., 2007a \cite{mypaperII}, Fig.~1) is to  
find galactic dark matter haloes to be very compact compared to field galaxies of
equivalent luminosity: an upper limit on the half mass radius is set at 50\,kpc,
when similar studies performed on field galaxies inferred half mass radius larger
than 200\,kpc. The mean total mass for the galaxy sample is found of order 0.2\,10$^{12}$\,M$_{\rm o}$.
These results are in good agreement with former galaxy-galaxy lensing studies through cluster
core based on \textsc{hst} data 
(Natarajan et~al., 1998 \cite{priya1}; Geiger \& Schneider, 1999 \cite{geigeramas}; 
Natarajan et~al., 2002a,b \cite{priya2,priya3}; Natarajan et~al., 2008 \cite{priya4}).

This observational results is interpreted within the tidal stripping scenario.
The theoretical expectation is that the global tidal field of a
massive, dense cluster potential well should be strong enough to
truncate the dark matter halos of galaxies that traverse the cluster
core (Avila-Reese et~al., 2005 \cite{avila05}; Ghigna et~al., 2000 \cite{ghigna00};
Bullock et~al., 2001 \cite{bullock}).
In other words, when a galaxy is falling into a cluster, it will experiment strong tidal 
forces that will strip their dark matter haloes, feeding the cluster dark matter halo itself.

\subsection{Comparison with N-body hydrodynamical simulations}
In order to get more insight into the tidal stripping scenario and to interpret the observational
results within a theoretical framework, I have been analyzing 
high resolution, N-body hydrodynamical simulations of two fiducial galaxy clusters
(Limousin et~al., 2008 \cite{mypaperIII}). These
simulations include dark matter and baryonic particles (Sommer-Larsen et~al., 2005 \cite{romeo3}).
The effect of tidal stripping on cluster galaxies hosted
in these dark matter subhalos as a function of cluster-centric radius is investigated.
To quantify the extent of the dark matter halos of cluster galaxies, we introduce the
half mass radius \rhalf as a diagnostic, and study its evolution with
projected cluster-centric distance \Rc as a function of redshift. We
find a well defined trend for \rhalfRc : the closer the galaxies are
to the center of the cluster, the smaller the half mass
radius. Interestingly, this trend is inferred in \emph{all} redshift
frames examined in this work ranging from $z$\,=\,0 to $z$\,=\,0.7.
At $z$\,=\,0, galaxy halos in the central regions of clusters are
found to be highly truncated, with the most compact half mass radius
of 10\,kpc (Fig.~1). We compare these results with galaxy-galaxy lensing probes
of \rhalf and find qualitative agreement. We make predictions concerning the evolution
of \rhalf with cluster-centric radius that will be possible to probe with future surveys using
space based telescopes such as \textsc{snap}, that combine wide field and high resolution imaging.

\begin{figure}
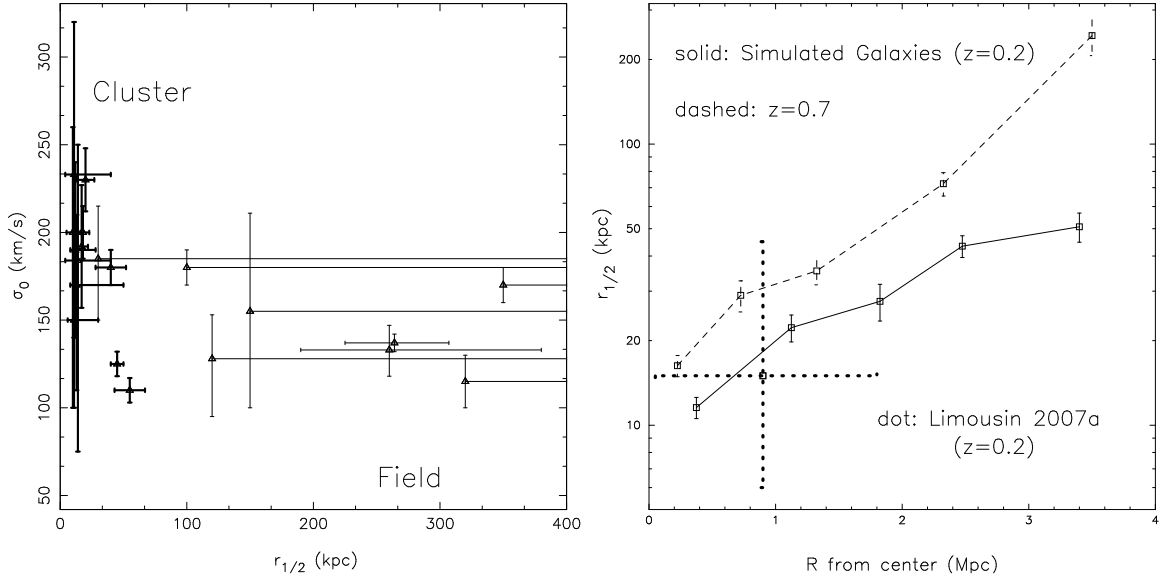

\psfig{figure=gglensingreview.ps,height=3.0in}
\psfig{figure=Moriondrhalf.ps,height=3.0in}
\label{gglensing_simu}
\caption{Left: Comparison between galaxy-galaxy lensing results on cluster galaxies
(thick black) and galaxy-galaxy lensing results on field galaxies (thin grey): cluster galaxies appear to be
significantly more compact than field galaxies.
References are given in Limousin et~al., 2007a.
Right: Evolution of the half mass radius as a function of cluster centric distance
for the simulated galaxies at $z=0.2$ (solid line) and $z=0.7$ (dashed line). 
Dotted data point correspond to the galaxy-galaxy lensing constraints derived in 
Limousin et~al., 2007a.
}
\end{figure}

\section{Combining Strong and Weak Gravitational Lensing in Abell~1689}
\subsection{Strong Lensing Analysis}
Deep \textsc{hst acs} images have provided an unprecedented wealth of arcs in
Abell~1689. 34 multiply imaged systems have been reported and used in Limousin et~al., 2007b \cite{mypaperIII}.
24 systems have been confirmed spectroscopically.
Using these observational constraints, I built a mass model using the \textsc{lenstool}
software (Jullo et~al., 2007 \cite{jullo07}).
Thanks to the large number of multiple images, the projected mass distribution is very well constrained,
at the percent level accuracy.
This mass model has been used by Stark et~al., (2007) \cite{stark07} to detect very high redshift
candidates beyond redshift 8, thanks to the natural amplification provided by the cluster.
I also report the finding of five strong galaxy-galaxy lenses located at $\sim
300$ kpc from the cluster center, i.e. \emph{outside} the critical region of the cluster.
These events have been used as an independent probe of the cluster potential in
Tu et~al., (2008) \cite{ring1689}.

\subsection{Weak Lensing Analysis}
Strong lensing allows us to constrain the projected mass distribution in the cluster core 
(typically 50`` from the cluster centre). To probe the cluster potential at larger
cluster-centric radii (up to 1\,000''), I have been using wide field multi-color data 
from the \textsc{cfh12k} camera.
From a background galaxy catalog, it is straightforward to construct a shear profile,
but a difficult issue is to reliably know who has been lensed by the cluster, since any 
contamination of the background galaxy catalog with non lensed cluster galaxies will dilute the shear
signal and bias the derived parameters (Fig.~2).
Using Bayesian photometric redshift, I have selected a background galaxy population.
The strong lensing and the weak lensing regimes are found to agree for the first time in this 
extensively studied cluster. The global concentration parameter is found to be around 8, 
a value which is rather high but compatible with $\Lambda$\textsc{cdm} predictions (Neto et~al., 2007
\cite{neto}). 
Former analysis based on Subaru data on the other hand have reported concentration parameters
larger than 20 (see, for example, Medezinsky et~al., 2007 \cite{elinor}), which is problematic
for the $\Lambda$\textsc{cdm} scenario. Since it is important to discriminate between these values,
we have been revisiting the weak lensing analysis from a mosaic of 16 \textsc{hst} pointings
(Dahle, Limousin et~al., in prep.). This independent analysis leads to a concentration parameter
around 10. We thus conclude that there is no concentration problem in Abell~1689.
There remain some discrepancies between the lensing and the \textsc{x}-ray masses. 
However, analysis of new very deep \textsc{chandra} data (see contribution by Signe
Riemer-S{\o}rensen) reveal a complicated structure which is likely not to be in hydrostatic
equilibrium and elongated along the line of sight, which could explain the mass disagreement.
Moreover, this elongation along the line of sight (which is clearly supported by the spectroscopy
of bright cluster members) can also explain the large Einstein radius observed in this cluster and
which have been recently claimed to be problematic for $\Lambda$\textsc{cdm} (Broadhurst \& Barkana,
2008 \cite{bb}).

\begin{figure}
\psfig{figure=Moriond_1689shear_1.ps,height=3.0in}
\psfig{figure=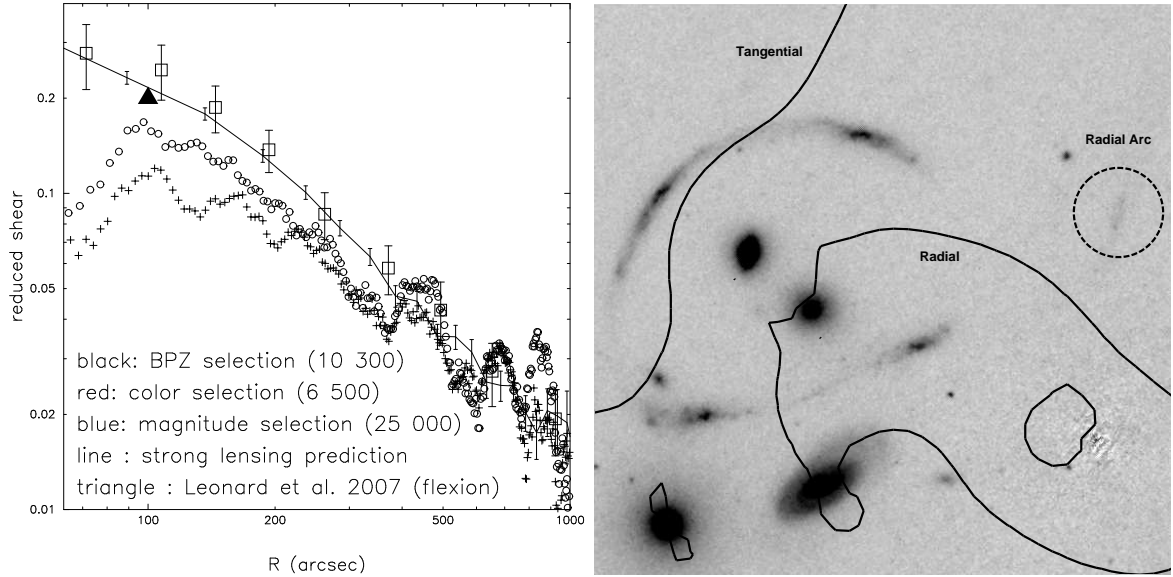,height=3.0in}
\label{amas}
\caption{Left: Comparison between shear profiles constructed using different rejection criteria to
select background lensed sources: the Bayesian photometric redshift based selection (squares),
a color selection (circles), and a magnitude cutoff (plus signs). The number of objects corresponding to
each rejection criterion is given in parentheses. The solid line corresponds to the strong-lensing
prediction, and the filled triangle comes from the study by Leonard
et~al., (2007).
Right: 'central ring' in Abell~1703, composed of four bright images. F775W image where the light from 
the cD galaxy
has been subtracted (note residuals). We plot the tangential and critical lines at the redshift of this
system.
}
\end{figure}

\section{Strong Lensing in Abell~1703: Constraints on the Slope of the Inner Mass Distribution}
Properties of dark matter haloes are probed observationally and numerically, and comparing both
approaches provide constraints on cosmological models.
When it comes to the inner part of galaxy cluster scale haloes, interaction between the baryonic and
the dark matter component is an important issue which is far to be understood.
Efforts are needed on the observational and numerical side in order to understand what is going
on in the centre of the most massive virialized structures of the Universe.
In a recent article (Limousin et~al., 2008 \cite{mypaperV}), we have been trying to measure the slope
of the inner dark matter distribution in Abell~1703, using strong lensing techniques.
Abell~1703 is a massive \textsc{x}-ray luminous galaxy cluster at $z=0.28$. The 
analysis is based on imaging data both from space
and ground in 8 bands, complemented with a spectroscopic survey.
Abell~1703 looks rather circular from the general shape of its multiply imaged systems and present a dominant
giant elliptical cD galaxy in its centre. This cluster exhibits a remarkable bright '\emph{central ring}'
formed by 4 bright images at $z_{\rm spec}=0.888$ located very close to the cD galaxy, providing observational constraints
that are potentially very interesting to probe the central mass distribution (Fig.~2).
The stellar contribution from the cD galaxy ($\sim$\,1.25\,10$^{12}$\,M$_{\rm o}$ within 7") is
accounted for in the parametric mass modelling, and the underlying
smooth dark matter component distribution is described using a generalized NFW profile parametrized with a central
logarithmic slope $\alpha$. We find that within the range where observational constraints are present
(from $\sim\,5"$ to $\sim\,50"$), the slope of the dark matter distribution in Abell~1703 is
equal to $1.09^{+0.05}_{-0.11}$ (3$\sigma$ confidence level).
The concentration parameter is equal to $c_{200} \sim 3.5$, and the scale radius is constrained to be larger than
the region where observational constraints are available. Within this radius, the 2D mass is equal
to M(50")\,=\,2.4\,10$^{14}$\,M$_{\rm o}$.
We cannot draw any conclusions on cosmological models at this point since we lack results from realistic
numerical simulations containing baryons to make a proper comparison.
Such comparison, when possible, may provide an interesting cosmological probe, in particular
when it comes to interactions between dark matter and baryons.

%All files (.tex .eps and .ps) should be put by the {\bf 15th of May 2008}
%on the upload website (http://www.apc.univ-paris7.fr/Uploads/incoming.html) with the "Target directory" = {\bf Moriond}
%for Yannick Giraud-Héraud
%({\bf ygh@apc.univ-paris7.fr}).\\

\section*{Acknowledgments}
I thank the organizers for providing me an opportunity to present my work.
The work reviewed in this article has been done thanks to the participation of various
friends and colleagues: \'A. El\'\i asd\'ottir,  B. Fort, E. Jullo, H. Dahle, 
H. Tu, J.-P. Kneib, J. Richard, J. Sommer-Larsen, O. Czoske, P. Natarajan, S. Bardeau \& T. Verdugo.
Marie Curie is acknowledged for the grant that covered my lodging expenses.
Funding from the Agence Nationale de la Recherche is acknowledged.
The Dark Cosmology Centre is founded by the Danish National Research Foundation.
\section*{References}
%\begin{thebibliography}{99}
%\bibliography{references}
\bibliography{limousin}

\begin{thebibliography}{10}

\bibitem{mypaperI}
M.~{Limousin}, J.-P. {Kneib}, and P.~{Natarajan} MNRAS, 2005

\bibitem{rix}
P.~{Schneider} and H.-W. {Rix}  ApJ, 1997.

\bibitem{olliphd}
O.~{Czoske}  PhD thesis, 2002

\bibitem{bardeau05}
S.~{Bardeau}, J.-P. {Kneib}, O.~{Czoske}, G.~{Soucail}, I.~{Smail},
  H.~{Ebeling}, and G.~P. {Smith} A\&A, 2005

\bibitem{deep2}
A.~L. {Coil} et~al. ApJ, 2004 

\bibitem{mypaperII}
M.~{Limousin}, J.~P. {Kneib}, S.~{Bardeau}, P.~{Natarajan}, O.~{Czoske},
  I.~{Smail}, H.~{Ebeling}, and G.~P. {Smith} A\&A, 2007a

\bibitem{priya1}
P.~{Natarajan}, J.-P. {Kneib}, I.~{Smail}, and R.~S. {Ellis} ApJ, 1998

\bibitem{geigeramas}
B.~{Geiger} and P.~{Schneider}
MNRAS, 1999

\bibitem{priya2}
P.~{Natarajan}, J.-P. {Kneib}, and I.~{Smail}
ApJ, 2002a

\bibitem{priya3}
P.~{Natarajan}, A.~{Loeb}, J.-P. {Kneib}, and I.~{Smail}
ApJ, 2002b

\bibitem{priya4}
P.~{Natarajan}, J.-P. {Kneib}, I.~{Smail}, T.~{Treu}, R.~{Ellis}, S.~{Moran},
  M.~{Limousin}, and O.~{Czoske}
ApJ, submitted

\bibitem{avila05}
V.~{Avila-Reese}, P.~{Col{\'{\i}}n}, S.~{Gottl{\"o}ber}, C.~{Firmani}, and
  C.~{Maulbetsch}
ApJ, 2005

\bibitem{ghigna00}
S.~{Ghigna}, B.~{Moore}, F.~{Governato}, G.~{Lake}, T.~{Quinn}, and
  J.~{Stadel}
ApJ, 2000

\bibitem{bullock}
J.~S. {Bullock}, T.~S. {Kolatt}, Y.~{Sigad}, R.~S. {Somerville}, A.~V.
  {Kravtsov}, A.~A. {Klypin}, J.~R. {Primack}, and A.~{Dekel}
MNRAS, 2001

\bibitem{mypaperIII}
M.~{Limousin}, J.~{Richard}, E.~{Jullo} et~al. 
ApJ, 2007b

\bibitem{romeo3}
J.~{Sommer-Larsen}, A.~D. {Romeo}, and L.~{Portinari}
MNRAS, 2005

\bibitem{jullo07}
E.~{Jullo}, J.-P. {Kneib}, M.~{Limousin}, {\'A}.~{El{\'{\i}}asd{\'o}ttir},
  P.~J. {Marshall}, and T.~{Verdugo}
New Journal of Physics, 2007

\bibitem{stark07}
D.~P. {Stark}, R.~S. {Ellis}, J.~{Richard}, J.-P. {Kneib}, G.~P. {Smith}, and
  M.~R. {Santos}
ApJ, 2007

\bibitem{ring1689}
H.~{Tu}, M.~{Limousin}, B.~{Fort}, C.~G. {Shu}, J.~F. {Sygnet}, E.~{Jullo},
  J.~P. {Kneib}, and J.~{Richard}
MNRAS, 2008

\bibitem{neto}
A.~F. {Neto}, L.~{Gao}, P.~{Bett}, S.~{Cole}, J.~F. {Navarro}, C.~S. {Frenk},
  S.~D.~M. {White}, V.~{Springel}, and A.~{Jenkins}
MNRAS, 2007

\bibitem{elinor}
E.~{Medezinski}, T.~{Broadhurst}, K.~{Umetsu}, D.~{Coe}, N.~{Ben{\'{\i}}tez},
  H.~{Ford}, Y.~{Rephaeli}, N.~{Arimoto}, and X.~{Kong}
ApJ, 2007

\bibitem{bb}
T.~{Broadhurst} and R.~{Barkana}
ApJ, Submitted

\bibitem{mypaperV}
M.~{Limousin}, J.~{Richard}, J.~P. {Kneib}, H.~{Brink}, R.~{Pello}, H.~{Tu},
  J.~{Sommer-Larsen}, E.~{Jullo}, E.~{Egami}, M.~J. {Michalowski},
  R.~{Cabanac}, and D.~P. {Stark}
A\&A, in press

\end{thebibliography}
%\bibitem{bd}C.D. Buchanan {\it et al}, \Journal{\PRD}{45}{4088}{1992}.

%\end{thebibliography}

\end{document}